\documentclass[final,twocolumn]{elsarticle}

\usepackage[english]{babel}
\usepackage[T1]{fontenc}
\usepackage[utf8]{inputenc}
\usepackage[draft,hidelinks=true]{hyperref}
\usepackage{amssymb,amsmath}
\usepackage{siunitx}
\usepackage{graphicx}
\usepackage{algorithmic}
\usepackage{algorithm}
\usepackage{tikz}
\usetikzlibrary{calc,trees,positioning,arrows,chains,shapes.geometric,%
    decorations.pathreplacing,decorations.pathmorphing,shapes,%
    matrix,shapes.symbols,scopes,arrows.meta}
\usepackage{tkz-euclide}

\journal{Computer Physics Communications}
\providecommand{\xb}{\mathbf{x}}

\providecommand{\Bb}{\mathbf{b}}

\begin{document}

\begin{frontmatter}

\title{CADISHI: Fast parallel calculation of particle-pair distance histograms
on CPUs and GPUs}

\author[mpcdf]{Klaus Reuter\corref{cor1}}
\ead{klaus.reuter@mpcdf.mpg.de}

\author[bio]{Jürgen K\"ofinger}
\ead{juergen.koefinger@biophys.mpg.de}


\cortext[cor1]{Corresponding author}

\address[mpcdf]{Max Planck Computing and Data Facility, Gießenbachstraße 2,
85748 Garching, Germany}

\address[bio]{Max Planck Institute of Biophysics, Max-von-Laue-Straße 3, 60438
Frankfurt, Germany}

\begin{abstract}

We report on the design, implementation, optimization, and performance of the
CADISHI software package, which calculates histograms of pair-distances of
ensembles of particles on CPUs and GPUs. These histograms represent 2-point
spatial correlation functions and are routinely calculated from simulations of soft
and condensed matter, where they are referred to as radial distribution
functions, and in the analysis of the spatial distributions of galaxies and
galaxy clusters. Although conceptually simple, the  calculation of radial
distribution functions via distance binning requires the evaluation of
$\mathcal{O}(N^2)$ particle-pair distances where $N$ is the number of particles
under consideration.
CADISHI provides fast parallel implementations of the distance histogram algorithm for
the CPU and the GPU, written in templated C++ and CUDA. Orthorhombic and general
triclinic periodic boxes are supported, in addition to the non-periodic case.
The CPU kernels feature cache-blocking, vectorization and thread-parallelization
to obtain high performance. The GPU kernels are tuned to exploit the memory and
processor features of current GPUs, demonstrating histogramming rates of up to a
factor 40 higher than on a high-end multi-core CPU. To enable high-throughput
analyses of molecular dynamics trajectories, the compute kernels are driven by
the Python-based CADISHI engine. It implements a producer-consumer data
processing pattern and thereby enables the complete utilization of all the CPU
and GPU resources available on a specific computer, independent of special
libraries such as MPI, covering commodity systems up to high-end HPC nodes.
Data input and output are performed efficiently via HDF5. 
In addition, our CPU and GPU kernels can be compiled into a standard C library
and used with any application, independent from the CADISHI engine or Python.
The CADISHI software is freely available under the MIT license.
\end{abstract}

\begin{keyword}
radial distribution function, pair-distance distribution function, two-point correlation
function, distance histogram, GPU, CUDA
\end{keyword}

\end{frontmatter}


{\bf PROGRAM SUMMARY}

\begin{small}
\noindent
{\em Program Title:} CADISHI                                  \\
{\em Licensing provisions:} MIT                               \\
{\em Programming language:} C++, CUDA, Python                 \\
{\em Nature of problem:}\\ 
Radial distribution functions are of fundamental importance in soft and
condensed matter physics and astrophysics. However, the calculation of distance
histograms scales quadratically with the particles number. To be able to analyze
large data sets, fast and efficient implementations of distance histogramming
are crucial.
\\
{\em Solution method:}\\ 
CADISHI provides parallel, highly optimized implementations of distance
histogramming.  On the CPU, high performance is achieved via an advanced cache
blocking scheme in combination with vectorization and threading.  On the GPU,
the problem is decomposed via a tiling scheme to exploit the GPU's massively
parallel architecture and hierarchy of global, constant and shared memory
efficiently, resulting in significant speedups compared to the CPU.  Moreover,
CADISHI exploits all the resources (GPUs, CPUs) available on a compute node in
parallel.
\\
{\em Additional comments including Restrictions and Unusual features (approx. 50-250 words):}\\
CADISHI implements the minimum image convention for orthorhombic and general
triclinic periodic boxes.  We provide Python interfaces and the option to
compile the kernels into a plain C library.
\end{small}


\section{Introduction}

Radial distribution functions link the structural and thermodynamic properties
of soft and condensed matter \cite{Hansen:13, McQuarrie:75}. Structurally, these particle pair correlation functions provide the probability of finding a
particle at a certain distance from another particle of a system.
Thermodynamically, these functions determine the equation of state
for systems with pair-wise interactions. In astronomy and astrophysics, these spatial two-point correlation functions are used to describe the distribution of galaxies or galaxy clusters in the universe \cite{Springel2005,Kerscher2000}.

In experiments on condensed matter, the Fourier transform of the radial
distribution function is measured by elastic scattering of x-rays or neutrons,
or light scattering in the case of microscopically sized particles like
colloids. Scattering intensities can be calculated accurately from radial
distribution functions using Debye's equation \cite{Debye:47}. This approach is
used, for example, to calculate small- and wide-angle x-ray scattering
intensities from molecular dynamics simulations of biological macromolecules in
solution \cite{Koefinger2013,CapriqornGithub}. 

Radial distribution functions are central to liquid state theory and facilitate
the interpretation of molecular simulations. For simple liquids, we can predict
phase transitions of liquid mixtures using radial distributions functions
obtained from the integral theory of Ornstein and Zernike \cite{Ornstein:14,
Hansen:13}.  In simulations, radial distribution functions help us to interpret
ordering effects and the resulting effective interactions between particles
\cite{Likos:00}. For complex systems, we usually lack feasible theoretical
approaches to calculate radial distribution functions. We thus estimate radial
distribution functions from molecular dynamics simulation (MD) trajectories and
Monte Carlo simulation ensembles by calculating histograms of particle
pair-distances \cite{FRENKEL200263}.

This task of calculating a radial distribution function scales with the number
of particles squared and is thus computationally challenging for large systems.
Large-scale parallel molecular dynamics simulations of soft and condensed matter
in explicit solvent can generate large amounts of trajectory data with hundreds
of thousands of frames with potentially millions of particles per frame
\cite{Vendruscolo2011}. Levine, Stone, and Kohlmeyer were the first to tackle
this challenge by taking advantage of the processing power of CUDA-enabled GPUs
\cite{Levine20113556}. Their software can be easily used via VMD \cite{VMD}, a
widely used program to set up, visualize, and analyze molecular dynamics
simulations. 
Extending their pioneering efforts, we provide here a novel software for the
"CAlculation of DIStance HIstograms" (CADISHI), which uses both CPUs and GPUs on
a single node to calculate radial distribution functions at very high
performance.  In addition to non periodic systems, orthorhombic and general
triclinic boxes are supported.

This paper is structured as follows.  Section \ref{sec:methods} briefly
introduces the mathematical background and discusses sequential and parallel
distance histogramming.  Section \ref{sec:implementation} details how
the distance histogram algorithms are efficiently implemented on the CPU and on
the GPU.  We report and discuss extensive benchmark results for both
kinds of processors in section \ref{sec:performance}.
Finally, section \ref{sec:summary} closes the paper with a summary.

\section{Methods}

\label{sec:methods}

To calculate a radial distribution function from an ensemble of particles, we
calculate all pair distances and collect them in a histogram. If the ensemble
stems from molecular simulations then we have to properly take into account the
boundary conditions. Commonly, we use periodic boundary conditions in
simulations and apply the minimum image convention. We distinguish two
scenarios: In the first scenario, we are interested in bulk properties and  we
calculate radial distribution functions up to half the minimum image distance,
e.g., half the box-length for a cubic box. In the second scenario, we simulate a
single macromolecule in a box as a model for a dilute system.  To calculate
scattering intensities (SAXS/WAXS), we have to cut out the macromolecule and a
sufficiently thick layer of solvent and effectively embed this system in
infinite solvent \cite{Koefinger2013}.  In this case, we calculate the radial
distribution function for the complete sub-system we have cut out, without
applying any periodic boundary conditions.

In the following, we first show how radial distribution functions are calculated
from histograms of pair-distances following the notation of Levine et
al. \cite{Levine20113556}. We then sketch the basic algorithm, recapitulate how
to take periodic boundary conditions for orthorhombic and triclinic boxes into
account, and discuss different methods for the implementation and
parallelization.

\subsection{Mathematical background}

The radial distribution function \cite{Hansen:13, McQuarrie:75} is defined as
\begin{equation}
    g(r) = \lim\limits_{dr \rightarrow 0}{\frac{p(r)}{4\pi(N_\mathrm{pairs}/V)r^2 dr}}.
\end{equation}
Here, $r$ is the distance between a pair of particles, $p(r)$ is the average
number of atom pairs found at a distance between $r$ and $r + dr$, $N_\mathrm{pairs}$
is the total number of unique atom pairs in the system, and V is the total
volume of the system.
For MD simulations, $p(r)$ is calculated from a finite number of trajectory
frames $N_\mathrm{frames}$ for all unique atom pairs indexed by $i,j$ as
\begin{equation}
    p(r) = \frac{1}{N_\mathrm{frames}} \sum_{k}^{N_\mathrm{frames}} \sum_{i, j (\neq i)}
    \delta(r - r_{ijk}) \text{,}
\end{equation}
where $r_{ijk}$ is the distance between particles $i$ and $j$ at frame $k$.
The $\delta$ function is replaced by a uniform histogram on a grid by
introducing
\begin{equation}
    p(r) = \frac{1}{N_\mathrm{frames}} \sum_{k}^{N_\mathrm{frames}} \sum_{i, j (\neq i)}
    \sum_{\kappa} d_{\kappa}(r, r_{ijk}) \text{.}
\end{equation}
Here, $\kappa$ is the histogram bin index. The value of a histogram bin is defined as
\begin{equation}
    d_{\kappa}(r, r_{i,j,k}) =
    \begin{cases}
        \Delta r^{-1} \; \text{if} \; r_\kappa \leq r, r_{ijk} < r_\kappa + \Delta r \\
        0 \; \text{else.}
    \end{cases}
\end{equation}
where $\Delta r$ is the width of a histogram bin,
and $r_\kappa = \kappa \Delta r$ is the lower bound of a histogram bin.

\subsection{Basic sequential particle pair distance histogram computation}

\begin{algorithm}[tb]
\begin{algorithmic}
\STATE  initialize histogram to zero
\FOR{$i=0$ to $N_1$ \COMMENT{loop over species 1}}
\FOR{$j=0$ to $N_2$ \COMMENT{loop over species 2}}
\STATE  compute distance vector $d\xb_{ij}$ between particle $i$ and particle
        $j$,
\STATE  apply minimum image convention to $d\xb_{ij}$ \COMMENT{for periodic
        systems only},
\STATE  compute distance
        $r_{ij} = (x_{ij}^2 + y_{ij}^2 + z_{ij}^2)^{1/2}$
\STATE  obtain bin index $\kappa_{ij} = (\texttt{int}) n_{bins} r_{ij} / r_{max}$
\STATE  increment histogram bin at index $\kappa_{ij}$
\ENDFOR
\ENDFOR
\end{algorithmic}
\caption{Two-species particle pair distance histogramming algorithm.}
\label{alg2}
\end{algorithm}
Algorithm \ref{alg2} sketches the basic sequential two-species distance
histogram computation, which is a common use case.
From a set of $N_1$ particles of species 1 and a set of $N_2$ particles of
species 2, the distance for each combination of two particles is evaluated,
rescaled to an integer index which is finally used to increment the bin counter.
In periodic systems, the minimum image convention is applied to the
distance first.  In total, distances between $N_1 \times N_2$ particle pairs
need to be binned.

Considering the single species distance histogram computation of a set with $N$
particles, the difference to algorithm \ref{alg2} is given by the constraint that we
perform only evaluations of unique pairs.  To this end, the inner loop in
algorithm \ref{alg2} is modified to start from $j = i + 1$, with $N = N_1 =
N_2$.  In this single species case, in total $N(N-1)/2$ particle pairs need to
be considered.

It is obvious that the floating-point computation intense part of the algorithm is given by
the distance computation. Further numerical costs are added by periodic
boundary conditions, which introduce the need for rounding operations and, in
the case of the general triclinic box, the need for multiple distance evaluations
between the different images.

On a side note, it seems tempting to avoid the numerically costly square root operation in the
distance calculation and to use a quadratically scaled histogram instead.  However,
in practice it turns out that the resulting larger histogram array spoils the
cache efficiency and leads in combination with the necessary postprocessing of
the histogram to inferior results compared to a high-performance implementation
of the direct Euclidean distance computation.

\subsection{Periodic boundary conditions}

Commonly, molecular dynamics simulations and Monte Carlo simulations of soft and
condensed matter apply periodic boundary conditions  (PBCs) to minimize surface effects,
which would otherwise cause artifacts.  Moreover, using triclinic box geometries corresponding to the truncated octahedron or the rhombododecahedron, for example, we can calculate radial
distribution functions for larger distances than if we used an orthorhombic box
with the same number of particles. The use of triclinic boxes can also increase
performance of simulations of single macromolecules by minimizing the number of
solvent particles we have to add.
In a periodic box the distance between two particles is given by the distance
between one particle in the central box and the nearest image of the other
particle. CADISHI implements this so-called the minimum image convention for
both, the orthorhombic and the general triclinic periodic box on both the CPU and
the GPU, as detailed on in the following.

Under the minimum image convention, the distance vector $d\xb^\prime$ between two points in a general periodic box is given by
\begin{equation}
d\xb^\prime = d\xb - \Bb \; \texttt{nint}\left( \Bb^{-1} d\xb \right)\, ,
\label{eq:orthorhombic}
\end{equation}
where $d\xb$ is the difference vector between the images of two particles in the
same box, $\Bb$ is the $3 \times 3$ matrix of box vectors, and \texttt{nint}
denotes rounding to the nearest integer. For further details, we refer to
Appendix B in the book by Tuckerman \cite{Tuckerman}.

An orthorhombic box has a rectangular basis and the basis vectors can have
different lengths. In this case, $\Bb$ and $\Bb^{-1}$ are diagonal which
simplifies the evaluation of equation (\ref{eq:orthorhombic}) in practice.

A general triclinic box has three basis vectors of different lengths which
intersect at arbitrary angles. Hence, an orthorhombic box is a special case of
the triclinic box.
Equation (\ref{eq:orthorhombic}) gives the minimum distance vector between two
images for distances up to half of the minimum width of the periodic box.  Note
that $\Bb$ is now non-diagonal.
To support larger distances, the distance vector from equation
\ref{eq:orthorhombic} needs to be shifted to the neighboring cells in order to
find the true minimum distance.
We refer the reader to the PhD thesis of Tsjerk Wassenaar for details
\cite{Wassenaar}.

\subsection{Parallel particle-pair distance histogram computation.}


In the case of distance histogramming, the distance calculations between
particle pairs are embarrassingly data parallel. The computational load is
easily balanced between processing units by distributing equally sized subsets
of the data, i.e.~the combinatorial set of all the relevant particle pairs.
Handling the bin updates and the computation of the final histogram correctly in
parallel turns out to be more intricate. Basically, one of the following
approaches is possible, as pointed out in Ref.~\cite{Levine20113556}.

First, all the processing units may update the bins of a single shared histogram
concurrently, requiring atomic hardware instructions or other mechanisms to
synchronize the individual memory updates.
Second, each processing unit may fill its own private histogram, followed by a
global reduction to sum up all the private partial histograms in order to obtain
the final histogram.
Third, a combination of both the previous approaches may be favorable, where
groups of processing units share a private histogram.
As will be detailed on in the following sections, the second option is suitable
for the CPU whereas the third option is well suited for the GPU.

\section{Implementation}
\label{sec:implementation}


The CPU and GPU kernels discussed in the following sections are implemented in
templated C++.  Templates avoid code duplication and allow, for example, to compile
executables for single and double precision coordinate input from the same
source code. Moreover, templates facilitate the generation of efficient code for all
supported cases, i.e., with or without periodic boxes, and with or without the
check if a distance falls within the maximum allowed value in the non-periodic case.
This is critical for performance because branches in inner loops are avoided
completely at runtime.  The distance computation including the minimum image
convention is implemented only once and used by both, CPU and GPU, via header
file inclusion.  We provide Python interfaces to the kernels.


Finally, to enable processing of large-scale MD simulation data, the CPU and GPU
kernels need to be driven efficiently to exploit all the compute resources
available on a computer.  To this end we have implemented a Python layer labeled
the CADISHI engine which is presented in section \ref{sec:CADISHI}.

\subsection{Histogram computation on CPUs}
\label{sec:histocpu}


An efficient implementation of the particle-pair distance histogram algorithm
needs to take advantage of  all levels of parallelism and caches of modern x86\_64 CPUs.
First, there are several physical cores per chip which typically support more
than one hardware thread each (simultaneous multithreading, called
hyper-threading for Intel CPUs).
Second, each core supports SIMD parallelism on vectors with a width of 128
(SSE2), 256 (AVX), or even 512 bits (AVX512), being able to operate on 4, 8, and 16
single precision numbers with a single instruction, respectively.
Moreover, to hide memory latencies, a cache hierarchy exists with individual
caches per physical core (L1 and L2), and caches shared between multiple cores
(L3).
Finally, in multi-socket machines, different chips may access the same physical
memory, however, at different latencies and bandwidths (NUMA).

To optimize for cache utilization, we implemented cache-blocked versions of the algorithms in addition to a direct implementation of the double-loop structure of algorithm \ref{alg2}.  The cache-blocked versions target the L2 cache, which has a size of 256 kB to 1 MB on modern CPUs and is therefore large enough to
hold for each thread a block of coordinates, an index buffer, and the partial
histogram.
\def\eps{0.05}
\begin{figure}[t]
\centering
\begin{tikzpicture}
    \tkzInit[xmax=3.2,ymax=3.2,xmin=0,ymin=0]
    \tkzGrid[sub,subxstep=.2,subystep=.2]]
    \draw[color=white, fill=white, fill opacity=0.75] (0.,0.) -- (0.,3.25) -- (3.25,3.25) -- cycle;
    \tkzAxeX[label=$j$]
    \tkzAxeY[label=$i$]
    \draw[color=red,semithick] (0,0) -- (3.2,3.2);
    \draw[color=cyan,semithick] (1.0 +\eps,1.0)
                              -- (1.0 +1.0-\eps,1.0+1.0-2.*\eps)
                              -- (1.0+1.0-\eps,1)
                              -- cycle;
    \draw[color=cyan,semithick] (1.6,1.15) node {1,1};
    \draw[color=blue,semithick] (2,1) rectangle (2. + 1. - \eps, 1. + 1. - \eps)
                                node[pos=.5] {2,1};
    \draw[decorate,decoration={brace,amplitude=4pt},xshift=-3pt,yshift=0pt]
            (0.,1.) -- (0.,2.0-\eps) node {};
    \draw (-0.5,1.55) node {$bs$};
    \draw[decorate,decoration={brace,amplitude=4pt},xshift=0pt,yshift=-3pt]
            (3.-\eps, 0.) -- (2., 0.) node {};
    \draw (2.5, -0.5) node {$bs$};
\end{tikzpicture}

\begin{tikzpicture}
    \tkzInit[xmax=3.4,ymax=2.2,xmin=0,ymin=0]
    \tkzGrid[sub,subxstep=.2,subystep=.2]]
    \tkzAxeX[label=$j$]
    \tkzAxeY[label=$i$]
    \draw[color=blue,semithick] (2,1) rectangle (2. + 1. - \eps, 1. + 1. - \eps)
                                node[pos=.5] {2,1};
    \draw[decorate,decoration={brace,amplitude=4pt},xshift=-3pt,yshift=0pt]
            (0.,1.) -- (0.,2.0-\eps) node {};
    \draw (-0.5,1.55) node {$bs$};
    \draw[decorate,decoration={brace,amplitude=4pt},xshift=0pt,yshift=-3pt]
            (3.-\eps, 0.) -- (2., 0.) node {};
    \draw (2.5, -0.5) node {$bs$};
\end{tikzpicture}

\caption{\label{fig:tiling} Loop tiling for the distance histogram calculation in the case of a single particle species (top) and in the case of two particle species (bottom).  For a single particles species, the tiles are
triangular directly below the diagonal and quadratic or rectangular elsewhere,
with edges of maximum length $bs$. For two particle species, the $i$ axis indicates the tile index of the first species and the $j$ axis indicates tile index of the second species.}
\end{figure}
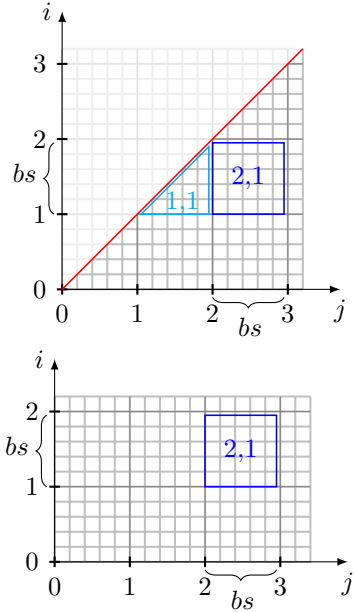

In our loop tiling, shown schematically in Fig.~\ref{fig:tiling}, the block
size $bs$ is defined as the length of an edge of a tile. The block size depends
on the histogram width $\texttt{n\_bins}$ and is given by the solution of
\begin{align}
\label{eq:blocksize}
\texttt{L2\_cache\_size} \nonumber - \texttt{reserve} + \texttt{extension} \nonumber \\
= \texttt{sizeof(uint32\_t)} \cdot bs^2 \nonumber\\
+ 2 \cdot \texttt{sizeof(coordinate\_tuple)} \cdot bs \nonumber \\
+ \texttt{sizeof(uint32\_t)} \cdot \texttt{n\_bins}\, ,
\end{align}
which is a simple quadratic equation.
On the left-hand side of Eq.~(\ref{eq:blocksize}) we define the amount of cache
in bytes we make available for cache blocking. This amount is determined by the
size of the cache \texttt{L2\_cache\_size}, determined at runtime once during
initialization, by the amount of of cache we reserve for other data
\texttt{reserve}, and by \texttt{extension}. Here, we set \texttt{reserve} to 16
kB. The value of \texttt{extension} is set to zero per default. For histograms
wider than 45k bins,  the value of \texttt{extension} is increased proportional
to the histogram width in order to avoid that the block size gets too small and
to enable the blocking scheme also for wide histograms. The transition at 45k
bins was determined by benchmarking.

The right-hand side Eq.~(\ref{eq:blocksize}) depends on the size
$\texttt{sizeof(uint32\_t)}$ of the cache array for the indices, the size of the
two sets of particle coordinates, $2\cdot \texttt{sizeof(coordinate\_tuple)}$,
and the width of the histogram, $\texttt{sizeof(uint32\_t)} \cdot
\texttt{n\_bins}$. As shown in the benchmark section below, the blocking scheme
allows to scale to large problem sizes without any performance degradation.  At
small problem sizes, the non-blocked versions are faster and a heuristic in the
code decides which kernel is to be called for a specific input.

The CPU kernels are threaded by means of OpenMP directives. Each thread works
on a subset of the combinatorial set of all relevant particle pairs and updates
its own private instance of the histogram.  The cache-blocked implementation
parallelizes over tiles, whereas the non-blocked version parallelizes the
double-loop directly using OpenMP directives.  For each thread, the final step
consists of the reduction of all the private partial histograms into the
complete histogram.

In addition to threading, SIMD parallelism (vectorization) is of key importance
to achieve good performance on modern CPU cores. Two implementation details turn out
to be essential for vectorization.
First, to adjust the memory alignment, the coordinate triples must be padded,
either implicitly via compiler-specific attributes or explicitly by extending the
triple by a fourth dummy element.
Second, incrementing a histogram bin immediately after each distance
calculation would access memory in a non-predictable fashion and
potentially lead to cache thrashing. Thus, the bin indices are stored temporarily in a
contiguous buffer, which is of size $bs^{2}$ in case of the kernels with cache
blocking. The bin updates are done from that buffer array, an
operation that is inherently non vectorizable due to its non contiguous memory
access pattern on the histogram array.
Note that we do not use intrinsics for portability reasons.  Hence, it is up to
the compiler to actually vectorize the code which may fail in some cases as shown
below.
The software Intel Amplifier XE was used to guide the optimization work.

For performance reasons, we use integers for bin counts and thus have to take
care to avoid integer overflows.  The thread-private histograms use unsigned 32
bit integers (\texttt{uint32\_t}).  Whenever a thread's number of processed
particle pairs approaches the upper limit of \texttt{uint32\_t}, the
thread-private histogram is added atomically to the global histogram
(``flushed'') and reset to zero.  Th global histogram uses unsigned 64 bit
integers (\texttt{uint64\_t}). Compared to a pure 32 bit implementation, the
performance penalty turns out to be marginal. This strategy is also applied in
the GPU implementation.

\subsection{Histogram computation on GPUs}

GPUs are not only successfully used to speed up simulations of molecular
dynamics \cite{Levine2011,Kutzner2015}, they are similarly well suited to
accelerate analysis tasks such as the pair-distance histogram calculation.
State-of-the art graphics processing units are comprised of several streaming
multiprocessors that have on the order of 10 to 100 cores each.  All the
multiprocessors have access to a cached single global memory on the GPU,
distinct from the host's main memory. We partly follow the lines of Levine
et.~al.~\cite{Levine20113556} for the implementation of a tiling scheme suitable
to obtain high performance.  Moreover, we extend their work, e.g.~with kernels
enabled to scale into the large bin number regime and with the support of
triclinic periodic boxes.

Our implementations are based on the NVIDIA CUDA programming model \cite{CUDA}
but the key points made are applicable to other platforms as well. In the spirit
of a heterogeneous programming model, kernels are launched from the host code to
perform computation on the GPU using numerous lightweight threads.  Logically,
CUDA organizes threads in thread blocks, and arranges the thread blocks on a
grid.  All threads from a thread block run on the same streaming multiprocessor,
grouped into so called warps of 32 threads that run simultaneously.  Threads of
the same thread block are able to communicate via a shared memory that can be
regarded as a user-managed cache.  Different thread blocks are independent.
When multiple threads read from the same memory address at the same time, GPU
constant memory with its associated constant memory cache is highly beneficial.
For further details, we refer the reader to the NVIDIA CUDA documentation
\cite{CUDA}.

With typically on the order of a thousand to a million of particles per MD
trajectory frame, the combinatorial set of all particle pairs contains
on the order of $10^{6}$ to $10^{12}$ elements.  Clearly, the large number of
independent distance calculations can be mapped to independent threads very well
in order to keep the numerous GPU cores busy.  However, the efficient handling
of the histogram bin updates is a major challenge.  This operation involves
some kind of synchronization between threads and is, moreover, characterized by
scattered memory accesses.
%


%

In general, the data transfer between host and GPU memory is a performance
critical aspect.  For the distance histogram computation, the cost of the data
transfer is insignificant for sufficiently large problem sizes due to the
computational complexity $\mathcal{O}(N^2)$, nevertheless we take several optimization steps.  To avoid the overhead from multiple individual transfers, the
complete multi-species coordinate set of an MD trajectory frame is prepared in a
contiguous memory area in pinned memory on the host and is then copied to a
contiguous area of GPU global memory in a single operation. The GPU kernels
then calculate the histograms for all combinations of particle species. Finally,
the resulting set of histograms is transferred from GPU global memory back to
CPU pinned memory in a single operation.  GPU memory is allocated at the first
kernel call only and reused at subsequent calls.

To obtain the optimum performance for all relevant input parameters and to allow
for cross validation, three kernel implementations of increasing complexity were
developed that are explained in detail below.
In the production code, the most suitable kernel is selected together with its optimum launch
parameters for a particular GPU model based on the particle numbers and on the histogram
width.  To this end, the implementation
internally provides heuristics covering recent GPU architectures.  The NVIDIA
Visual Profiler was used to guide the optimization work.

In the following sections, we present the GPU histogram kernel implementations in order of increasing complexity.  


\subsubsection{A basic GPU kernel}

A straight-forward approach to implement the pair-particle distance histogram
calculation using the CUDA framework is to map the double-loop structure of
algorithm \ref{alg2} to a two-dimensional grid of thread blocks such that
each thread works on an individual particle pair.  Bin increments are naively
implemented by performing atomic updates of a single shared histogram in global
memory.  The overhead of this approach can be reduced by choosing the actual
CUDA grid smaller than the total grid spanned by the number of particle
coordinates. Doing so we let each thread work on several particle pairs by looping
over the full coordinate set using the CUDA total grid size as offset.  Consequently, global memory accesses are coalesced automatically.

On older GPUs (pre Kepler), the binning rate could be increased by cloning the
histogram bins in global memory and reducing them in a final step, which lowers the
 access frequency of individual bins and therefore the collision rate.  On Kepler
and more recent GPUs we used during the final development of this work,
we find that the new fast atomic operations in global memory do not make such
cloning necessary any more \cite[Kepler tuning guide]{CUDA}.

We refer to this implementation as the \emph{simple} kernel.  It is
characterized by a rather flat performance profile independent of the number of
bins. In the scope of this work it is only used to perform correctness checks
and as the starting point for more complex kernels.  It is outperformed
substantially by the two implementations presented in the following, which are
actually intended for production use.

\subsubsection{Improving performance by coordinate tiling in constant memory}
\label{sec:globalknl}

To speed up the simple GPU kernel, a tiling scheme is introduced in the spirit
of a cache blocking technique aimed at a reduction of the number of accesses to global
memory.  We first make use of the GPU's constant memory segment in order to
accelerate the memory access to the particle coordinate data.  Limited in size
to 64 kB, constant memory has an associated fast on-chip cache that delivers a
value as quickly as if it was read directly from a register, provided that all
the threads in the warp access the same address.  However, data can be copied to
constant memory only from the host code.

In our implementation, the inner loop of algorithm \ref{alg2} is mapped to a
one-dimensional CUDA grid, covering the second coordinate set stored in global
memory. The outer loop is written explicitly inside the kernels.  It iterates
over a tile of the first coordinate set that is stored in constant memory.
Hence, each thread reads a coordinate tuple from global memory in a coalesced
fashion and performs the distance binning for all the particles stored in the
constant memory tile. The latter is the \emph{same} for all the threads at
each loop iteration such that we exploit the fast constant memory cache.  We added to the
host code an additional loop which handles the copying of coordinate data to
constant memory before launching the kernel.  For single precision
data, a tile in the 64 kB of constant memory comprises about 5300 coordinate
tuples. For each such tile a kernel launch has to be performed.

To optimize the performance for frames with several different particle species
and different particle numbers, we minimize the number of kernel launches for
two-species histograms.We first sort the particle coordinate sets by increasing
particle number. As a consequence, the set for the second species located in
global memory has typically more members than the first one located in constant
memory. This arrangement reduces the number of necessary kernel launches.

We label this implementation the \emph{global memory} kernel because it keeps the
histogram in global memory. It turns out to be the fastest kernel when going to
larger histogram bin numbers, as we will demonstrate below. In addition, it is
an important intermediate step towards optimizing the kernel further, as will be
done in the following by introducing a tiling scheme for the histogram in shared
memory.

\subsubsection{Improving performance by histogram tiling in shared memory}
\label{sec:sharedknl}

The global memory kernel presented in the previous section optimizes the
coordinate tuple accesses but performs the atomic bin updates in
comparably slow global memory.  The key step to increase the binning performance
is to introduce private partial histograms in shared memory.  Compared to
pre-Maxwell GPUs, the performance benefits greatly from the fast shared memory
atomic operations that became available with Maxwell chips \cite[Maxwell tuning
guide]{CUDA}.

In general, one shared memory buffer can be allocated per CUDA thread block.  It
can be accessed by all the threads in the block at near register speed.
However, shared memory is a scarce resource and limited to a maximum of $48$ kB
per thread block on most of the GPUs relevant to the present work.  Up to
now, only the Volta V100 GPU can be configured to provide $96$ kB of shared
memory to a thread block, which significantly improves performance of this implementation
\cite[Volta tuning guide]{CUDA}. Hence, 32 bit partial histograms in shared
memory can be at most $12288$ ($\times 2$ on the V100) bins wide. Consequently,
to allow the kernels to process wider histograms, the binning range must be
tiled, requiring multiple sweeps through all the particle pairs, increasing the
run time proportionally.  Before the lifetime of a thread block ends, the
private partial 32-bit histogram in shared memory is added atomically to the
64-bit histogram in global memory.

\begin{figure*}[t]
\centering
\begin{tikzpicture}[align=center, node distance=1.cm and 2.cm]
    \providecommand\MinHeight{0.8cm}
    \providecommand\MinWidth{1.8cm}
    \providecommand\OuterSep{0.0cm}
    \providecommand\Shorten{0.15cm}

    \tikzstyle{CPUMEM}=[draw, rounded corners, minimum height=1.5*\MinHeight, minimum width=\MinWidth,
                        fill=blue!50, anchor=south west, outer sep=\OuterSep]

    \tikzstyle{CPUMEM2}=[draw, rounded corners, minimum height=\MinHeight, minimum width=\MinWidth,
                        fill=blue!50, anchor=south west, outer sep=\OuterSep]

    \tikzstyle{GPUMEM}=[draw, rounded corners, minimum height=1.5*\MinHeight, minimum width=\MinWidth,
                        fill=red!50, anchor=south west, outer sep=\OuterSep]

    \tikzstyle{GPUMEM2}=[draw, rounded corners, minimum height=\MinHeight, minimum width=\MinWidth,
                        fill=red!50, anchor=south west, outer sep=\OuterSep]

    \tikzstyle{GPUCONST}=[draw, rounded corners, minimum height=0.8*\MinHeight, minimum width=\MinWidth,
                          fill=gray!50, anchor=south west, outer sep=\OuterSep]

    \tikzstyle{GPUSHARED}=[draw, rounded corners, minimum height=0.5*\MinHeight, minimum width=\MinWidth,
                           fill=violet!50, anchor=south west, outer sep=\OuterSep]

    \tikzstyle{GPUBLOCK}=[draw, rectangle, minimum height=\MinHeight, minimum width=\MinWidth,
                          fill=orange!50, anchor=south west, outer sep=\OuterSep]

    \tikzstyle{ARROW}=[shorten >=\Shorten, shorten <=\Shorten, -{Latex[length=2.75mm, width=1.5mm]}]

    \node[GPUCONST] (gpuconst) at (0,0) {subset of\\coord\_1\\(constant)};
    \node[CPUMEM, left = of gpuconst] (cpusel1) {coord\_1};

    \node[GPUMEM, below = of gpuconst] (gpusel2) {coord\_2\\(global)};

    \node[CPUMEM, left = of gpusel2] (cpusel2) {coord\_2};

    \node[CPUMEM, left = of gpusel2] (cpusel2) {coord\_2};
    \node[GPUMEM2, below = of gpusel2] (gpuhisto) {histogram\\(global)};
    \node[CPUMEM2, left = of gpuhisto] (cpuhisto) {histogram};

    \node[GPUBLOCK, right = of gpusel2]     (threadblock2) {
        thread\_block\_2\\
        \tikz\node[GPUSHARED] {partial histogram (shared)};
    };
    \node[GPUBLOCK, below = 0.5cm of threadblock2]     (threadblock3) {
        thread\_block\_N\\
        \tikz\node[GPUSHARED] {partial histogram (shared)};
    };
    \node[GPUBLOCK, above = 0.5cm of threadblock2]     (threadblock1) {
        thread\_block\_1\\
        \tikz\node[GPUSHARED] {partial histogram (shared)};
    };
    \node[below = 0.125cm of threadblock2]     (threadblockdots) {
        \ldots
    };

    \draw[ARROW] (cpusel1) to (gpuconst);
    \draw[ARROW] (cpusel2) to (gpusel2);
    \draw[ARROW] (gpuhisto) to (cpuhisto);

    \draw[ARROW] (gpuconst.east) to (threadblock1.west);
    \draw[ARROW] (gpusel2.north east) to (threadblock1.west);

    \draw[ARROW] (gpuconst.east) to (threadblock2.west);
    \draw[ARROW] (gpusel2.east) to (threadblock2.west);

    \draw[ARROW] (gpuconst.east) to (threadblock3.west);
    \draw[ARROW] (gpusel2.south east) to (threadblock3.west);

    \draw[ARROW] (threadblock1.west) to (gpuhisto.east);
    \draw[ARROW] (threadblock2.west) to (gpuhisto.east);
    \draw[ARROW] (threadblock3.west) to (gpuhisto.east);

    \draw[dashed] (-1.0,-4.25) -- (-1.0,2.25);
    \node[above = 0.5cm of cpusel1] (cpulabel) {CPU};
    \node[right = 5.0cm of cpulabel] {GPU};
\end{tikzpicture}
\caption{Tiling scheme employed for the parallel decomposition
on the GPU, including the use of the memory hierarchy.  From CPU memory (left,
blue), a set of coordinate data from the first species is copied to GPU constant
memory (constant, grey) before each kernel launch.  The coordinate data of the
second species is copied initially once to global GPU memory (global, red). The loop
over that second set of coordinates is implemented via a 1d grid of thread
blocks, such that each thread block operates on a different subset of coord\_2.
Each thread block internally loops over the first species, accessing the
identical  element at a time from each thread and thereby exploiting the fast
constant memory cache. Each thread block updates its private partial histogram
in shared memory (shared, violet).  Before a thread block terminates, the private
partial histogram is added atomically to the histogram in global GPU memory,
which is finally copied back to CPU memory.
}
\label{fig:gpu}
\end{figure*}
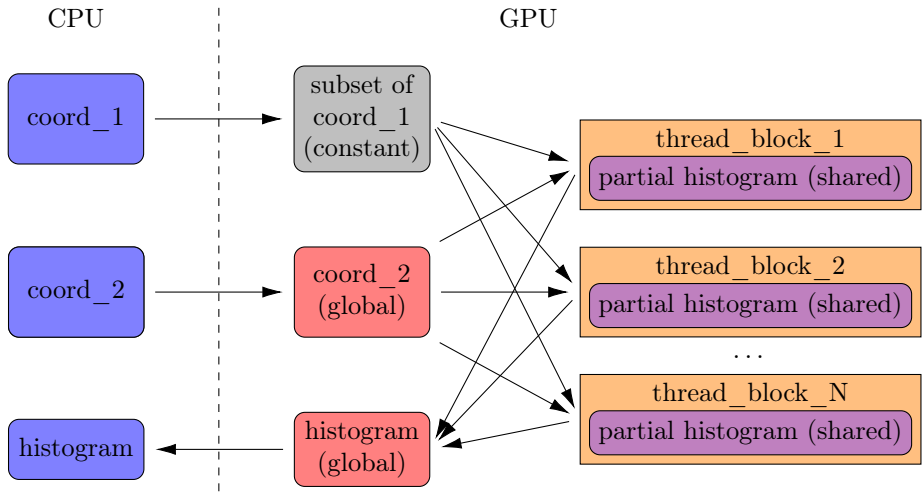
Figure \ref{fig:gpu} shows a schematic of our shared-memory-based GPU
implementation, providing a detailed explanation of the memory access in the
caption.

We label this implementation the \emph{shared memory} kernel.  It is by far the
fastest kernel at comparably small bin numbers, when only one or few sweeps
through the particle pairs is necessary.  For any requested number of bins, the
implementation determines the number of sweeps from the available shared memory.
The shared memory histograms are then tiled to have (about) the same size for
all the sweeps.

\subsubsection{GPU kernel optimization}

We determined the optimum block sizes of both, the global memory and the shared memory kernels
for the 1d CUDA grid from benchmark runs of sufficiently large problem sets, which saturate the GPUs.  These optimal size are chosen automatically by a simple heuristic. For Maxwell and newer GPUs, a block size of 512 threads is chosen, whereas on older (Kepler) GPUs it is beneficial to use larger blocks of
1024 threads. A user may override these defaults when calling the kernels.

In case of the shared memory kernel, we determine the size of the tiles for the partial
histograms by the number of sweeps necessary to compute a
requested histogram width. The number of sweeps is determined by the maximum amount of
shared memory available per thread block.
In general, as little shared memory as possible should be used in order to keep
the occupancy of the streaming multiprocessors sufficiently high, primarily to
hide global memory latencies.
The best performance is not necessarily achieved at an
occupancy of 100\%, which refers to the maximum number of threads a GPU is
able to keep active. For some compute-bound kernels the instruction-level
parallelism is able to use the GPU very well at occupancies smaller than one
\cite{Volkov2010}.
As shown in section \ref{sec:gpuperf} the distance histogram kernels fall into
this class and are able to saturate the GPUs at occupancies down to 25\%.

So far, we only discussed the two-species computation.
As pointed out before, for the single-species case the only difference is given
by the start index of the inner loop, such that duplicate and
same-particle evaluations are avoided.  In the GPU kernels, an \emph{if} branch is used to
determine from the thread index within the CUDA grid if the thread shall
evaluate a particle pair.  Note that for sufficiently large problem
sizes only a small fraction of the CUDA thread blocks are actually affected by
this if branch, very similar to the diagonal blocks required for the CPU cache
blocking (see fig.~\ref{fig:tiling}). If the condition is true for all the
threads of a warp, no overhead is introduced. If it is true for only some of
the threads, the branches are serialized, i.e., some threads of the warp stay
idle while the other threads perform their computations in parallel.

Intentionally, the present implementation does not split large sets of particle
pairs onto several GPUs.  Rather, motivated by realistic application scenarios
that require the processing of numerous frames, individual frames
are processed completely on a single GPU, allowing to exploit trivial frame
parallelism by using several GPUs.

\subsection{CADISHI parallel engine}
\label{sec:CADISHI}

Next, we discuss the CADISHI engine, which enables users to exploit all the resources (CPUs, GPUs) available on a compute node simultaneously. Such an efficient use of resource is especially useful for the parallel analysis of long MD trajectories with many frames.

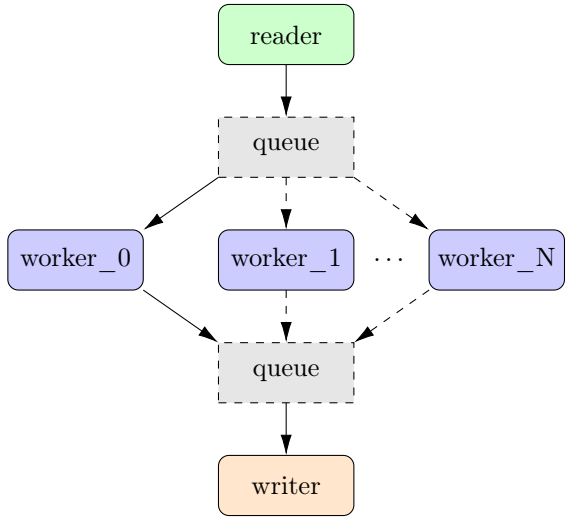
\begin{figure}[t]  
\centering
\begin{tikzpicture}[align=center, node distance=1.5cm]
    \providecommand\MinHeight{0.8cm}
    \providecommand\MinWidth{1.8cm}
    \providecommand\OuterSep{0.0cm}
    \providecommand\Shorten{0.0cm}
    \tikzstyle{COMMON} = [rectangle, draw, text centered, minimum height=\MinHeight, minimum width=\MinWidth, outer sep=\OuterSep]
    \tikzstyle{reader} = [COMMON, fill=green!20, rounded corners]
    \tikzstyle{worker} = [COMMON, fill=blue!20, rounded corners]
    \tikzstyle{writer} = [COMMON, fill=orange!20, rounded corners]
    \tikzstyle{queue} = [COMMON, dashed, fill=gray!20]
    \tikzstyle{ARR1} = [shorten >=\Shorten,
                        shorten <=\Shorten,
                        -{Latex[length=2.75mm, width=1.5mm]}]

    \node[reader]                       (reader) at (0,0) {reader};
    \node[queue, below of=reader]       (queuei) {queue};
    \node[worker, below of=queuei]      (worker1) {worker\_1};
    \node[worker, left=1cm of worker1]  (worker0) {worker\_0};
    \node[worker, right=1cm of worker1] (workerN) {worker\_N};
    \node[queue, below of=worker1]      (queueo) {queue};
    \node[writer, below of=queueo]      (writer) {writer};
    \draw[ARR1] (reader.south) -- (queuei.north);
    \draw[ARR1] (queuei.south west) -- (worker0.north east);
    \draw[ARR1] (worker0.south east) -- (queueo.north west);
    \draw[ARR1, dashed] (queuei.south) -- (worker1.north);
    \draw[ARR1, dashed] (worker1.south) -- (queueo.north);
    \draw[ARR1, dashed] (queuei.south east) -- (workerN.north west);
    \draw[ARR1, dashed] (workerN.south west) -- (queueo.north east);
    \draw[ARR1] (queueo.south) -- (writer.north);

    \node[right = 0.135cm of worker1] (dots) {\ldots};
\end{tikzpicture}
\caption{Schematic of the CADISHI trajectory data processing framework. Multiple
processes (reader, writer, workers) that communicate and synchronize via queues
are used to implement node-level parallelism. The workers either use GPU or CPU
resources.}
\label{fig:CADISHI}
\end{figure}
The concept of a data processing pipeline serves as the design principal, as shown schematically in figure \ref{fig:CADISHI}. Frames are
provided and buffered in a queue by a reader process.  To ensure
high performance, the input trajectory is read from HDF5 files \cite{HDF5}.  The
frames are picked up and processed in parallel by multiple worker processes,
each computing all the histograms for a particular frame using the CPU or GPU
kernels. The results are put into a second queue from which a writer process
fetches the histograms, averages them optionally, and saves them to HDF5.
To enable users to import MD simulation data easily, we provide a conversion
tool, which uses a generic reader \cite{MDA1,MDA2}.

This design offers a high degree of modularity and flexibility for current and
future methodical extensions, e.g., the integration into more complex analysis
workflows using the CAPRIQORN package \cite{CapriqornGithub, Koefinger2013}.  We
use the Python programming language to implement CADISHI. In particular, we use
Python's \texttt{multiprocessing} module, which is part of the standard library
\cite{PSTL} and enables node-level parallelism out of the box on virtually any
platform.  The implementation does not use nor require a third-party dependency
such as the message passing interface (MPI) for distributed-memory parallelism.
The cost of the inter-process communication of the atom coordinates and the
histograms is negligible compared to the $N^2$ complexity of the computational
problem.  The workers release the global interpreter lock of Python explicitly
when calling the compiled CPU and GPU kernels such that the inter-process
communication continues to run smoothly during the computation.

A useful configuration on a typical two-socket two-GPU compute node could be as
follows. Two CPU workers are used, each running the previously discussed
thread-parallel CPU histogram code on an individual CPU socket.  In addition,
two GPU workers are used, each of them running the GPU histogram code on an
individual GPU. It is important to reserve a physical CPU core for each, the
reader, the writer, and the GPU workers, in order to guarantee quick data
transfer and avoid IO becoming the bottleneck. CADISHI picks appropriate core
numbers and also handles the process pinning automatically.

\section{Performance benchmarks}
\label{sec:performance}

We performed extensive performance benchmarks
and investigate the binning rates of the CPU and GPU kernels individually for
input data of various sizes. The input data was generated by
putting particles at pseudo-random coordinates into a unit box.
To profile the CPU and GPU codes, we used a driver program to supply the
coordinate data, to launch the kernels, to determine the time-to-solution, and
to calculates the binning rate in billion atom pairs per second (bapps).  All
computations discussed below were performed in single precision, which is the
relevant use case when MD simulation data is processed.  In general, double
precision runs turn out to be between a factor of $1.3$ to $2$ slower on CPUs
and non-consumer GPUs, depending on the problem size and the presence of a
periodic box.
In all cases, we measured the time to solution, which includes memory transfers to and
from the GPU and overhead from GPU kernel launches.

In the second part, we show results for the node-level performance obtained on a compute
node with two GPUs, running the CADISHI engine on a practically
relevant data set from an MD simulation.

\subsection{CPU performance}
\label{sec:cpuperf}

The CPU kernel was profiled on a shared memory machine with two Intel Xeon
Platinum 8164 processors \cite{Skylake}, providing 26 physical cores and 52
hardware threads each.  The base clock of a core is 2 GHz, and the cores support
the AVX512 instruction set.

Benchmark results in this section are exclusively based
on the binary generated by the Intel compiler \texttt{icc}
compiler in version 2017 using the optimization flags \texttt{"-fast -xHost
-qopt-zmm-usage=high -qopenmp"}.  Indeed, the Intel compiler manages to generate
AVX512 code for the inner loop with the distance computation, covering all the
possible box cases. The instruction set in use was checked at runtime by reading out hardware
counters for floating point instructions via the Linux \texttt{perf} tool.

For comparison, we compiled kernels using the GNU \texttt{g++} compiler in
version 7.2 with the optimization flags \texttt{"-O3 -march=native -ffast-math
-funroll-loops -fopenmp"} applied. In doing so, vectorized AVX512 code is
generated by \texttt{g++} for the inner loop of algorithm \ref{alg2} in the case
without periodic boundary conditions.  In the cases of orthorhombic and
triclinic boxes, rounding operations and operations to determine the minimum
prevent the vectorization by the GNU compiler.  Compared to the binary generated
by \texttt{icc}, we find that the binary from \texttt{g++} runs virtually at the
same speed for the vectorized case without any box, whereas the cases with a
periodic box run significantly slower due to the lacking vectorization.

As will be seen in the next section, the GPUs perform much better in general,
which mitigates the drawback of having the boxed computation not vectorized well
on the CPU with the widely used GNU compiler.

\begin{figure}[t]
\centering
\includegraphics[width=1.0\columnwidth]{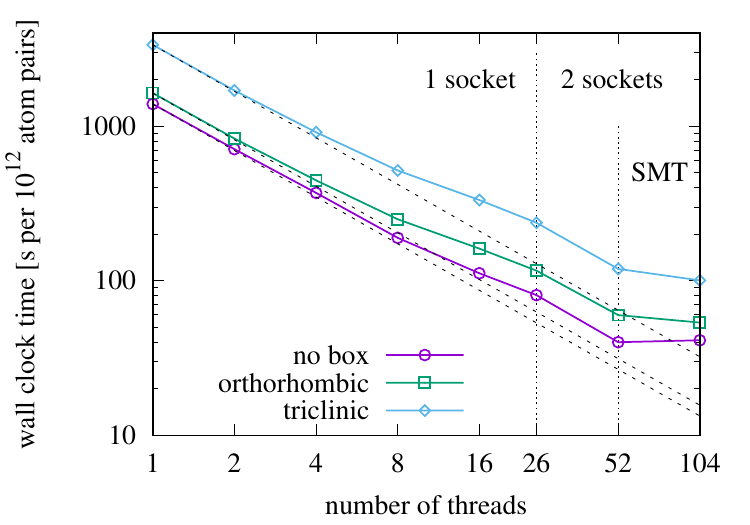}
\caption{Wall clock time of the histogram calculation as a function of the
number of OpenMP threads. Calculations were performed on the Skylake CPU for a
fixed problem size of $1\text{M} \times 1\text{M}$ particle pairs and
$10\text{k}$ bins, with cache-blocking enabled. The vertical dotted lines
indicate the transition from 1 to 2 sockets, and, moreover, the transition into
the simultaneous multithreading regime.}
\label{fig:openmp}
\end{figure}
Figure \ref{fig:openmp} shows a scan in the number of OpenMP threads for a fixed
problem size of $1\text{M} \times 1\text{M}$ particle pairs and $10\text{k}$
bins, covering the three possible box cases.  Initially, the number of threads
is increased from 1 up to all the 26 physical cores on a single socket. While
the scaling is ideal at the beginning, the curve starts to deviate from the
ideal scaling when the socket gets increasingly filled. Since the kernel is of
complexity $O(N^2)$ and therefore compute and not memory limited, this effect is
likely to be caused by the dynamic clocking of the vector units, which reduces
the frequencies as more and more cores are used in order to limit the heat
dissipation.  Memory accesses are of minor importance, in particular due to the
cache blocking optimization.
Going from 1 to 2 sockets, the scaling is ideal.  Enabling hyperthreading in
addition shows no benefit in the case without PBCs and only marginal benefit in the
other cases, indicating that the CPU pipelines are already used quite well. The
scaling is rather similar for the three cases.  The run times clearly indicate
the cost associated with the orthorhombic and the triclinic box, in particular.
For all the following investigations on the CPU, we use 26 threads on a single
socket, in order to provide a 1:1 baseline for the comparison with current GPU
models done in the following section.

\begin{figure}[t]
\centering
\includegraphics[width=1.0\columnwidth]{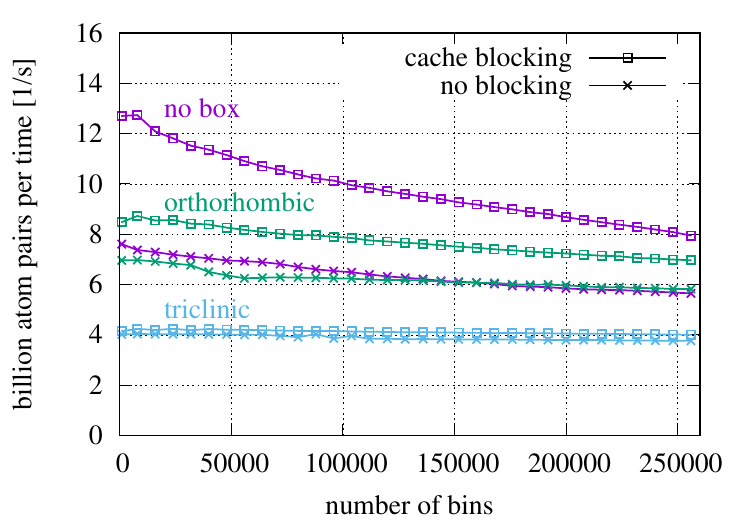}
\caption{Histogramming rate as a function of the number of bins. Calculations were performed on the Skylake CPU for a fixed problem size of $500\text{k} \times 500\text{k}$ atom pairs. We compare
the implementations with and without cache blocking for the cases without PBCs and with PBCs using orthorhombic and triclinic unit cells. }
\label{fig:cpu_scan_bins}
\end{figure}
Figure \ref{fig:cpu_scan_bins} shows a scan in the number of histogram bins on
the Skylake processor, comparing the implementations with and without the cache
blocking scheme for a fixed problem size of $500\text{k} \times 500\text{k}$
particle pairs and 10k bins, with and without periodic boxes.
At histogram widths of up to about 10k bins, the implementation with cache
blocking achieves the highest binning rate of about 13 bapps in the
case without PBCs. In the following, the binning rate decreases to about 10 bapps at
100k bins, around and beyond which a linear decrease is observed.
The reason for the initial steep decrease is that, on each core, the
thread-private histogram occupies an increasingly larger fraction of the L2
cache, as the number of bins is increased. Therefore, the coordinate blocks are
chosen accordingly smaller, decreasing the efficiency of the blocking scheme,
cf.~Eq.~\ref{eq:blocksize}.  Note that the cache blocking scheme was not
designed to optimize for scans in the histogram width but rather for large
problem sizes as will be discussed next.  Moreover, in practice many
applications require only moderate histogram widths which lie within the regime
of highest performance.
In comparison, the implementation without cache blocking is significantly
slower, for small bin numbers by about one third in the case without box.
The relative advantage from the cache blocking decreases when going to the
orthorhombic case and virtually vanishes for the triclinic case which is caused
by the increasing arithmetic intensity making the memory accesses less
important.

\begin{figure}[t]
\centering
\includegraphics[width=1.0\columnwidth]{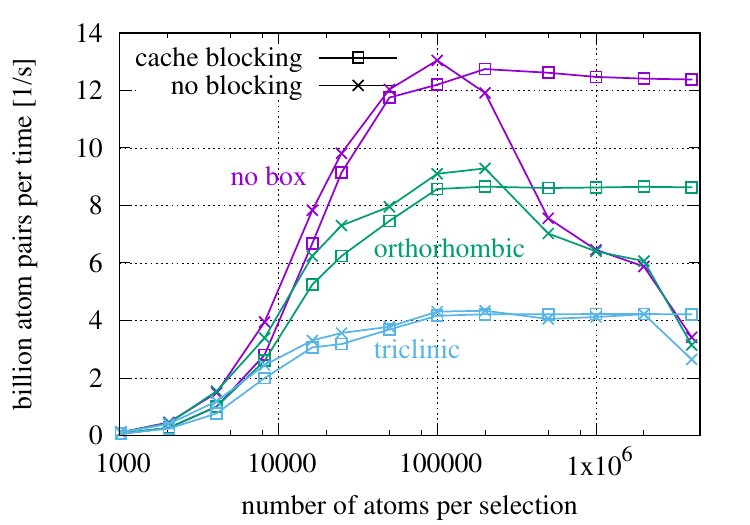}
\caption{Histogramming rate as a function of the problem size. Calculations were performed on the Skylake CPU for fixed histogram
width of 10k. We compare the cache-blocked with the non-blocked
implementation. In addition to the default case without PBCs, we show results for PBCs using  orthorhombic and
triclinic boxes.}
\label{fig:cpu_scan_size_boxes}
\end{figure}
Figure \ref{fig:cpu_scan_size_boxes} shows scans in the problem size while
keeping a fixed histogram width of 10k bins.  Results for the implementations
with and without cache blocking are shown.  In addition to cases without
PBCs, performance data for the kernels handling periodic boxes is
included.
We first discuss the scans without periodic box.  Up to a problem size of about
$100\text{k} \times 100\text{k}$ atom pairs, the code without cache blocking
turns out to be faster than the variant with blocking, clearly indicating some
overhead of the blocking scheme.  Going beyond that problem size, the
performance of the code with cache blocking is nearly constant, whereas the
performance without blocking drops severely by more than two thirds in the range
under consideration.
Turning towards the cases with periodic boxes, it is observed that the binning
rate is clearly slower compared to the case without periodic boxes.
Comparing the plateaus, the performance goes down to about two thirds for the
orthorhombic case and down to about one third for the triclinic case.
The overhead is partially caused by the \texttt{nearbyint()} function which is
used to perform the rounding during the application of the minimum image
convention.  In microbenchmarks, the \texttt{nearbyint()} function turned out to
be faster than the \texttt{round()} function by about $10\%$, which is likely
due to the fact that it does not raise the \emph{Inexact} exception on the CPU.
For the triclinic box, a three-fold loop over all neighbouring boxes is executed
in addition to find the minimum image for the general case, causing the
additional overhead.


\subsection{GPU performance}
\label{sec:gpuperf}

The CUDA code was compiled with the general optimization flags \texttt{"-O3
-use\_fast\_math"} and was in addition adapted for the GPU architecture under
consideration, e.g., by applying the flags \texttt{"--generate-code
arch=compute\_70, code=compute\_70"} for the Volta GPU. For the host code, GCC
and the same flags as previously were used.

\begin{table*}
\caption{Overview on the GPUs and host systems used for the performance
benchmarks. For more detailed specifications, we refer to the references
\cite{GTX1080,P100,V100,Minsky,DGX1V}. Note that the system with GTX1080 GPUs is
also used for benchmark runs based on real MD simulation data,
cf.~Sec.~\ref{sec:cadperf}.}
\centering
\begin{tabular}{lrrr}
\hline GPU & ($2 \times$) GTX1080 & ($4 \times$) P100 & ($8 \times$) V100 \\
\hline
fp32 peak [TFLOPS] & 8.87 & 10.6 & 15.7 \\
mem-bandw.~[GB/s] & 320 & 732 & 900 \\
Bus & PCIe 3.0  & NVLink & PCIe 3.0 \\
\hline
CPU & Intel Haswell        & IBM POWER8+     & Intel Broadwell \\
    & 2 $\times$ E5-2680v3 & 2 packages      & 2 $\times$ E5-2698v4 \\
Cores (Threads) & 2 $\times$ 12 ($\times$ 2) & 2 $\times$ 10 ($\times$ 8)  & 2 $\times$ 20 ($\times$ 2) \\
\hline \end{tabular}
\label{tab:gpus}
\end{table*}

We profiled the GPU kernel on several state-of-the-art hardware platforms as
shown in table \ref{tab:gpus}. We present results for NVIDIA V100 \cite{V100},
P100 \cite{P100}, and GTX1080 \cite{GTX1080} GPUs. The V100 GPU is part of an
NVIDIA DGX-1 system \cite{DGX1V}, whereas the P100 GPU is part of an IBM POWER8+
system \cite{Minsky}. In the DGX-1 system, the V100 GPUs are connected to the
host CPUs via PCIe 3.0.  In the POWER8+ system, the P100 GPUs are connected via
NVLink which is about a factor 3 faster in host-device bandwidth than PCIe 3.0.
Note that the GTX1080 GPU was designed for entertainment
applications, lacks ECC memory, and has only very few units enabled for double
precision operations.

Below we compare the performance of both implementations of interest.
First, we investigate the \emph{shared memory} kernel which keeps partial
histograms in shared memory, potentially requiring several sweeps through all
the particle pairs when going to larger bin numbers (see section
\ref{sec:sharedknl}).  In addition we profile the \emph{global memory} kernel,
which updates a single histogram in global memory (see section
\ref{sec:globalknl}) and turns out to be of advantage for larger bin numbers.

\begin{figure}[t]
\centering
\includegraphics[width=1.0\columnwidth]{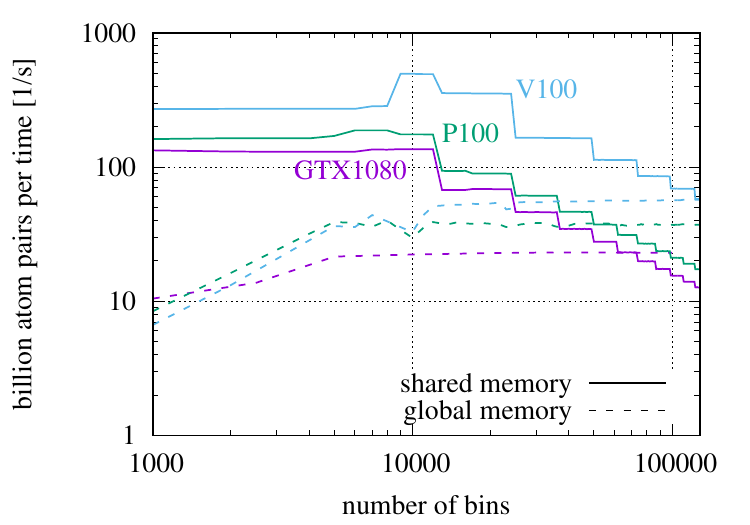}
\caption{Scan in the number of histogram bins for a problem size kept fixed at
$4\text{M} \times 4\text{M}$ atom pairs without any periodic box, comparing the
implementations using shared and global memory for the histogram updates on the
three GPUs under consideration.}
\label{fig:gpu_scan_size}
\end{figure}

We present scans in the histogram width in figure \ref{fig:gpu_scan_size}
with the problem size kept fixed at $4\text{M} \times 4\text{M}$ atom pairs,
which is large enough to saturate the GPUs (see below).
Initially, the performance curves for the shared memory kernels are constant at
rather high levels.  Remarkably, for the V100 GPU we find a plateau of highest
performance between 9k and 12k bins, where the curve peaks around 9k bins at a
binning rate of 495 bapps.
Following that initial range of highest performance, the binning rate decreases
in steps, with a width determined by the size of of the shared memory on the
GPU. Due to the limited memory, multiple sweeps are required for larger bin
numbers. Here, the V100 GPU has twice the step width due to its twice as large
shared memory of 96 kB per thread block.
The performance profile of the consumer-grade GTX1080 GPU is very similar to
that of the enterprise-grade P100 GPU, both featuring the Pascal microarchitecture.
The P100 GPU is somewhat faster, likely due to its higher internal memory
bandwidth.
In contrast, the performance curves of the global memory kernel rise initially
and reach plateaus at 12k bins. For larger bin numbers, the possibility of
collisions during bin updates in global memory is low.  Again the V100 GPU is
the fastest, followed by the P100 and GTX1080 devices.
The break-even point for the global memory kernels is around 60k bins for the
Pascal GPUs and at 128k bins for the V100 GPU. This information is used by the
implementation for a simple heuristic to decide which kernel is to be called for
a given bin number.

The achieved occupancy on the GPUs is nearly 100\% for histogram widths up to 4k
bins (8k on the V100), when the CUDA grid is chosen to launch 512 threads per
block for which we observe the best performance.
As confirmed using the \texttt{nvprof} tool the occupancy decreases down to
25\% as the histogram width in shared memory is increased to the maximum
possible value of about 12k (24k for the V100) bins.
%
%
The scans shown in figure \ref{fig:gpu_scan_size} indicate that the performance
of the shared memory kernel is not at all affected by the varying occupancy
which is due to their compute-bound characteristics and instruction-level
parallelism \cite{Volkov2010}.
Rather the number of sweeps is decisive as seen clearly from the staircase-type
curves.


\begin{figure}[t]
\centering
\includegraphics[width=1.0\columnwidth]{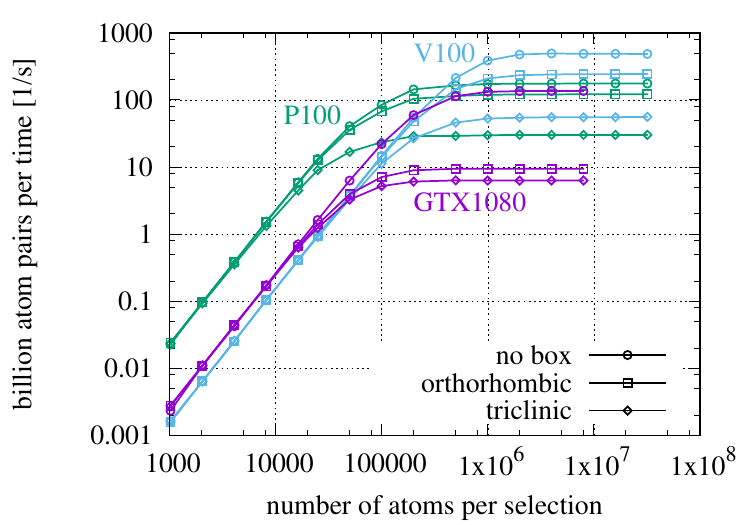}
\caption{Scan in the problem size for a fixed histogram width of 10k bins,
comparing the implementations with and without the periodic boxes on the three
GPUs under consideration.}
\label{fig:gpu_scan_size_boxes}
\end{figure}
Figure \ref{fig:gpu_scan_size_boxes} shows scans in the problem size for cases
with and without periodic boxes, while the histogram width is kept fixed at
10k bins.
Following the initial steep rise of the performance curves, it can be seen that
there are at least $10^{10}$ atom pairs required for the Pascal-based GPUs and
at least $10^{12}$ atom pairs for the V100 GPU to reach performance saturation
which is indicated by a flat-top in all cases.
For the P100 GPU, the effect from the fast NVLink interconnect is clearly
observed for small problem sizes when the GPUs are not saturated and the costs
of memory transfer and kernel launches are significant.  In particular, the
curves of the P100 device are shifted towards the left compared to the GTX1080
and the V100 GPUs which are connected via the relatively slower PCIe 3 bus to
the host CPUs.
The curves for the periodic boxes already saturate the GPUs for somewhat smaller
problem sizes and level off at lower binning rates, which is both due to the
higher arithmetic intensity.
Interestingly, on both the non-consumer-grade P100 and V100 GPUs and compared to
the results from the GTX1080 GPU, the performance of the periodic box cases is
much closer to the cases without PBCs.  For example, in the case of the orthorhombic box, the
binning rate with box is 0.69 (0.48) of the binning rate without box on the P100
(V100) GPU, whereas it is only 0.07 on the GTX1080 device.
The reason is likely to be linked to the fact that a significant part of the
Pascal processor's arithmetic capabilities are disabled on the consumer-grade
GTX1080, in particular a large fraction of the double-precision units.  We
speculate that this might apply as well to the round operation required by the
periodic boxes.
%

\subsection{Performance comparison between CPU and GPU}

\begin{table*}
\caption{Comparison between the performance on the Skylake chip with 26
cores and the performance on the three GPUs under consideration for a
problem size of $4\text{M} \times 4\text{M}$ atom pairs, 10k bins, without and
with periodic boxes. For the GPUs, the performance relative to the CPU is
given.}
\centering
\begin{tabular}{l|rr|rr|rr}
\hline
{}        & \multicolumn{2}{c|}{no box} & \multicolumn{2}{c|}{orthorhombic box} & \multicolumn{2}{c}{triclinic box} \\
processor & bapps [$s^{-1}$] & rel. & bapps [$s^{-1}$] & rel. & bapps [$s^{-1}$] & rel. \\
\hline
Skylake & 12.38 & 1.00 & 8.63 & 1.00 & 4.22 & 1.00 \\
GTX1080 & 135.66 & 10.96 & 9.44 & 1.09 & 6.28 & 1.49 \\
P100 & 175.04 & 14.14 & 120.89 & 14.01 & 30.11 & 7.14 \\
V100 & 494.45 & 39.95 & 239.34 & 27.74 & 55.25 & 13.10 \\
\hline
\end{tabular}
\label{tab:cpu_vs_gpu}
\end{table*}
%
%
Table \ref{tab:cpu_vs_gpu} gives a direct comparison between the performance on
the CPU and on the GPUs under consideration, based on data from the previously
discussed runs. A large problem size with $4\text{M} \times 4\text{M}$ atom
pairs was selected at which the performance on both types of processors is
saturated, and a histogram width of 10k bins, where the highest binning rates
are seen on both types of processors.  Absolute and relative performance numbers
are compared for the cases without and with periodic boxes.
In all cases the GPUs are significantly faster than the CPU.  While the consumer
grade GTX1080 is competitive with the professional Pascal model P100 in the case
without periodic box, the GTX1080 is significantly outperformed in the
orthorhombic and triclinic cases, presumably linked to disabled floating point
rounding capabilities.
The Volta V100 GPU is by far the fastest, beating the 26 core Skylake by a
factor of up to $\sim 40$ in the case without any box.

Note that our implementations are optimized to be most efficient for small and
moderately large bin numbers up to 24k. In many application scenarios it is
possible and sufficient to choose the number of bins, i.e., the desired numerical
resolution of the one-dimensional radial distribution function, to lie within that range of
highest performance.
Finally, we point out that the CPU code is faster for small problem
sizes, below selections of about 100k atoms, which is due to the overhead
induced by the heterogeneous programming model and hardware of the GPU.

\subsection{CADISHI single-node application performance}
\label{sec:cadperf}

This section reports on the CADISHI application performance based on actual MD
simulation data measured on a standard 2-socket server equipped with 2 GPUs.

The data set under consideration comprises 2000 frames with about 280000 particles each from an MD simulation of
\emph{F$_1$-ATPase}. Simulations were performed using NAMD \cite{NAMD}.  With 9 chemical species, 36 partial histograms have to be
evaluated for all the possible intra- and inter-species combinations for each
frame, where the individual numbers of the particle-pairs are highly different,
ranging from $\sim 10^2$ to $\sim 10^{10}$, which is in particular challenging
for the GPU implementation.  No periodic box was considered.  The
compressed HDF5 trajectory file has a size of 9 GB.  A resolution of 8000
histogram bins was chosen, going up to a maximum radius of
$\SI{300}{\angstrom}$.  We summed partial
histograms at intervals of 100 frames and wrote the summed-up histograms to the disk, leading to a compressed
HDF5 output file of 18 MB in size.
The performance benchmarks were run on a HPC cluster with two Haswell CPUs
[E5-2680 v3, with 12 (24) physical cores (hardware threads) each] and two
consumer grade GTX1080 GPUs per node, and a shared GPFS file system on which the
IO was performed.

\begin{table*}
\caption{Overview on the single-node CADISHI application performance achieved
for the F$_1$-ATPase dataset with 2000 frames, measured on the 2-socket Haswell node
with 2 GTX1080 GPUs.  The time was taken until all the partial histograms
were written to disk, i.e., the total time to solution is given, whereas the
performance in bapps is reported per compute worker and does not include
buffering and IO time.  The use of simultaneous multithreading is indicated by
an asterisk ($^{*}$).  Three runs per setup were performed and averaged.}
\centering
\begin{tabular}{l|rr|rr|r}
\hline
setup   & \multicolumn{2}{c|}{workers} & \multicolumn{2}{c|}{bapps [$s^{-1}$]} & {time [s]} \\
{}      & CPU (threads) & GPU & CPU & GPU & {} \\
\hline
C1      & 1 (44$^{*}$)    & 0      & 8.2   & 0        & 9594.6 \\
C2      & 2 (22$^{*}$)    & 0      & 4.2   & 0        & 9422.9 \\
G2      & 0               & 2      & 0     & 117.1    &  348.8 \\
G2C1    & 1 (20)          & 2      & 6.6   & 118.1    &  341.7 \\
G2C2    & 2 (10)          & 2      & 3.3   & 116.0    &  351.3 \\
\hline
\end{tabular}
\label{tab:f1atpase}
\end{table*}

Table \ref{tab:f1atpase} gives performance numbers for a selection of setups
found to perform well.  Note that for the reader, writer, and any (potentially
present) GPU workers one physical core was reserved each, and that for the
setups with CPUs involved all the remaining physical cores were used.  The
waiting time for new work packages is about 10 ms for all the setups shown.
The plain CPU runs C1 and C2 processed the complete dataset in less than 3
hours.  Here, using two workers on separate NUMA domains turns out to be
slightly faster than using only one worker on both the domains.  Simultaneous
multithreading was enabled for the sake of a small speedup
(cf.~Fig.~\ref{fig:openmp}).
On the other hand, the plain GPU run G2 processed the 2000 frames in slightly
below 6 minutes.  Relative to the run C2 the speedup is 27.0, demonstrating that
the GPU has a significant advantage not only with synthetic benchmark data but
also with real MD simulation data.

It seems tempting to perform hybrid runs to further speed up the GPU runs,
however, our experiments indicate that keeping all the CPU cores busy in
addition to the GPUs does pay off only marginally, if at all.
For such hybrid runs we find that the number of threads per CPU worker must be
not more than the number of available physical cores. For larger numbers of
threads, hyperthreading clogs the threaded multiprocessing queues between the
reader, the writer, and the worker processes.
%
Only the case G2C1 with a single CPU worker is slightly faster than the plain
GPU case G2.  For hybrid runs, the imbalance in processing speed between the CPU
and the GPU leads to the situation that the CPU worker is processing a final
work package while the GPU workers have already finished.  This effect is still
significant in the present example with 2000 frames but will become less
important for very large frame numbers.  Moreover, a run-time system for
task-based parallelism may be helpful to mitigate such situations, see
e.g.~\cite{StarPU}.

\section{Summary and conclusions}
\label{sec:summary}

The CADISHI software achieves very high performance on both, CPUs and GPUs. The
kernels for both types of processors can be driven by the Python-based CADISHI
engine to enable high-throughput analysis of MD trajectories.  CADISHI
implements a producer-consumer model and thereby allows for the complete
utilization of all the CPU and GPU resources available on a specific computer,
independent of special libraries such as MPI, covering commodity systems up to
high-end HPC nodes. CADISHI enables the analysis of trajectories with many
thousands of frames in a minimum amount of time.  Processing 2000 frames with
280000 particles each of trajectory data from an F$_1$-ATPase simulation was
demonstrated to run in less than 6 minutes on a standard two-socket compute node
with two consumer-grade GPUs.

To achieve high performance on the CPU, we proposed a cache tiling scheme
tailored to fit the L2 cache size of a CPU core, OpenMP SIMD directives in
combination with a linear index buffer to help the compiler generate vectorized
code, and thread parallelism over tiles using classical OpenMP directives.  In
our test, the implementation performs and scales well up to a full shared
memory node consisting of two 26-core Intel Skylake processors.

Compared to running the optimized CPU code on a 26-core Intel Skylake processor,
we find that our optimized GPU code achieves a speedup of up to 40 on an NVIDIA
V100 GPU for a case without any periodic box. For orthorhombic and triclinic
periodic boxes the speedup is 28 and 14, respectively. The consumer-grade
GTX1080 GPU is competitive with the professional models, in particular for cases
without a periodic box.

Here, we confirmed the observation of Levine et al.~\cite{Levine20113556} that
the use of constant memory to cache coordinate data is the key to achieve high
performance, even though the maximum available constant memory per GPU did not
increase over the GPU generations. In contrast to Levine et al., we use scarce
shared memory exclusively for storing histogram data and not for actively
caching coordinate data. We rather rely on the GPU's native hardware caches.
Levine at al.~use shared memory to implement overflow protection, which we
handle differently via constant memory tiles of known size in combination with a
CUDA grid directly mapping the loop over the second particle species. The
histogram updates in both, global and shared memory, have seen significant
improvement due to the introduction of fast atomic instructions with recent GPU
generations \cite{CUDA}. Moreover, the 96 kB of shared memory per streaming
multiprocessor of the V100 GPU accelerates the computation for medium and large
bin numbers. This hardware feature was previously unavailable.

Importantly, we complement the work of Levine et al. \cite{Levine20113556} by
providing a high-performance CPU implementation featuring vectorization and
parallelization, the support for the triclinic box, template-based support for
single and double precision and for runtime checks of the distances to fit
within the maximum binning range, and the CADISHI parallel engine for node-level
parallelism, leveraging unprecedented possibilities of large-scale MD data
analysis. We provide Python interfaces and the option to compile the kernels
into a plain C library.

The CADISHI software package presented in this paper is available free of charge
in source code under the permissive MIT license at \cite{CADISHIGithub}. It can
be used together with the CAPRIQORN software package \cite{CapriqornGithub,
Koefinger2013} to calculate SAXS/WAXS scattering intensities from molecular
dynamics trajectories.

\section*{Acknowledgements}

We thank Prof.~Gerhard Hummer, Max Linke, and Dr.~Markus Rampp for fruitful
discussions. We thank Prof.~Kei-ichi Okazaki for providing an initial NAMD setup
for F$_1$-ATPase.  We acknowledge financial support by the Max Planck Society.

\bibliography{cadishi}

\end{document}